\documentclass{article}[11pt]
\usepackage[utf8]{inputenc} 
\usepackage{microtype}
\usepackage{amssymb,amsmath,amsthm}
\usepackage{csquotes}
\usepackage{caption}
\usepackage{subcaption}
\usepackage{graphicx}
\usepackage{rotating}
\usepackage{float}
\usepackage{natbib}
\usepackage{color}
\usepackage{mathtools}
\usepackage{appendix}
\usepackage{multirow}
\usepackage{siunitx}
\usepackage{blindtext}
\usepackage{rotating}[figuresleft]
\usepackage{authblk}

\newcommand{\absdiv}[1]{%
  \par\addvspace{.5\baselineskip}
  \noindent\textbf{#1}\quad\ignorespaces
}

\title{Using meta-analytic priors to incorporate external information for study evaluation}
\begin{footnotesize}
\author[1]{Thilo Welz}
\author[1]{Eric Knop,}
\author[2]{ Frank Konietschke,}
\author[2]{ Jan-Hendrik B. Hardenberg,}
\author[1,4]{ Markus Pauly,}
\author[3]{ Christian Röver}
\affil[1]{Department of Statistics, TU Dortmund University, Fraunhofer Str. 2-4, 44227 Dortmund, Germany}
\affil[2]{Institut für Biometrie und Klinische Epidemiologie, Charité - Universitätsmedizin Berlin, Charitéplatz 1, 10117 Berlin, Germany}
\affil[3]{Universitätsmedizin Göttingen, Institut für Medizinische Statistik, Humboldtallee 32, 37073 Göttingen, Germany}
\affil[4]{UA Ruhr, Research Center Trustworthy Data Science and Security, 44227, Dortmund, Germany}
\end{footnotesize}
\date{December 2022}
\begin{document}
\maketitle

\begin{abstract} 

\absdiv{Background} The COVID-19 pandemic has had a profound impact on health, everyday life and economics around the world. An important complication that can arise in connection with a COVID-19 infection is acute kidney injury. A recent observational cohort study of COVID-19 patients treated at multiple sites of a tertiary care center in Berlin, Germany identified risk factors for the development of (severe) acute kidney injury. Since inferring results from a single study can be tricky, we validate these findings and potentially adjust results by including external information from other studies on acute kidney injury and COVID-19.

\absdiv{Methods} We synthesize the results of the main study with other trials via a Bayesian meta-analysis. The external information is used to construct a predictive distribution and to derive posterior estimates for the study of interest. We focus on various important potential risk factors for acute kidney injury development such as mechanical ventilation, use of vasopressors, hypertension, obesity, diabetes, gender and smoking.

\absdiv{Results} Our results show that depending on the degree of heterogeneity in the data the estimated effect sizes may be refined considerably with inclusion of external data. Our findings confirm that mechanical ventilation and use of vasopressors are important risk factors for the development of acute kidney injury in COVID-19 patients. Hypertension also appears to be a risk factor that should not be ignored. Shrinkage weights depended to a large extent on the estimated heterogeneity in the model.

\absdiv{Conclusions} Our work shows how external information can be used to adjust the results from a primary study, using a Bayesian meta-analytic approach. How much information is borrowed from external studies will depend on the degree of heterogeneity present in the model.


\end{abstract}

\section{Introduction}
\label{sec:Intro}

Over the past years the global COVID-19 pandemic, caused by the SARS-CoV-2 virus, has had far reaching and damaging effects on society, both in a health and economic sense \citep{padhan2021economics}. One important health complication that can arise in connection with COVID-19 is acute kidney injury (AKI) \citep{batlle2020acute}. Reported rates of AKI suggest the kidney could be the second most affected organ in patients with COVID-19 \citep{gabarre2020acute,hardenberg2021critical}. However, the underlying mechanisms are not yet fully explored \citep{siew2020covid}. A recent observational cohort study of COVID-19 patients treated at multiple sites of a tertiary care center in Berlin, Germany observed the development of severe AKI in 70 out of 223 patients \citep{hardenberg2021critical}. Patients with severe AKI were older, mostly male and also had more comorbidities as well as excess mortality \cite{hardenberg2021critical}.

Inferring results from a single study however can be tricky, especially if the sample size is small. Researchers may be faced with one or several samples that have been gathered without necessarily being powered to achieve a certain pre-specified power. The overall aim for such studies may be rather exploratory in nature. This is the present context for a recent study on AKI in COVID-19 patients \cite{hardenberg2021critical}.  Here, a tempting approach is to include external information from other studies. This is also known as borrowing of strength \citep{rover2020dynamically}. A natural way to facilitate this is the application of Bayesian methods \citep{schmidli2014robust}. These allow to synthesize external studies via meta-analysis and to derive a predictive distribution for the main effect of a new study. We refer to this as the meta-analytic-predictive (MAP) prior \citep{mckinney2021developing,schmidli2014robust}. The MAP prior can be used as a prior for the new study in order to derive a posterior distribution via Bayes' theorem. This is equivalent to a meta-analytic-combined (MAC) approach, where all studies (including the new one) are meta-analyzed and a shrinkage estimate is subsequently calculated for the new study \citep{rover2019model}. This shrinkage estimate will be a weighted average between the meta-analytic estimate of the overall effect and the study-specific estimate. Whether the shrinkage estimate lies closer to the study-specific estimate or the meta-analytic estimate of the overall effect depends to a large degree on the heterogeneity present in the data. As highlighted in \cite{rover2021bounds}, it is even possible to calculate bounds for the \enquote{shrinkage weights}, i.e. the contribution of the study of focus to its own shrinkage estimate. We follow a MAC approach, in order to validate and potentially adjust the results in \cite{hardenberg2021critical}. We do this by incorporating study results from \cite{10.3389/fmed.2021.719472}, who conducted a systematic review and meta-analysis of (potential) risk factors for AKI in adult patients with COVID-19. Whereas \cite{hardenberg2021critical} mainly considered longitudinal data and temporal associations with disease course, we focus on baseline characteristics, due to a lack of comparable studies with longitudinal data.

In the following Section \ref{sec:methods}, we first review the standard normal-normal hierarchical model (NNHM) for meta-analysis in Subsection \ref{nnhm}. This is followed by a description of posterior distributions in Subsection \ref{posteriors} and a description of the shrinkage estimates of study-specific means in Subsection \ref{shrinkage}. In Section \ref{sec:aki-covid19} we introduce research on the occurrence of acute kidney injury and possible contributing factors in COVID-19 patients. We apply the described Bayesian methodology on data regarding acute kidney injury in COVID-19 patients from \cite{hardenberg2021critical} and \cite{10.3389/fmed.2021.719472} and present the results. We close with a discussion and an outlook in Section \ref{sec:discussion}.

\section{Methods}
\label{sec:methods}

\subsection{The normal-normal hierarchical model}\label{nnhm}
We briefly describe the standard NNHM, which can be used to address a wide range of problems and refer to \cite{borenstein2021introduction} for more details. In the NNHM, study-level effects are modeled using normal distributions, while heterogeneity is modeled as an additive normal variance component on a second hierarchy level. Given $k$ studies with their respective effects $y_i$ and (assumed known) sampling variances $\sigma_i^2$, with $i=1,\ldots,k$, the goal is to infer an overall effect $\mu$. The $y_i$ are estimates of $\mu$, which are assumed to be normally distributed around the $i$th study's true effect $\theta_i$: $y_i \sim \mathcal{N}(\theta_i,\sigma_i^2)$. In random-effects meta-analysis the study level effects $\theta_i$ are assumed to vary normally around the overall mean $\mu$ with heterogeneity variance $\tau^2$, i.e. $\theta_i \sim \mathcal{N}(\mu,\tau^2)$. The marginal formulation of this model is
\begin{equation}
    y_i \sim \mathcal{N}(\mu,\sigma_i^2 + \tau^2),
\end{equation}
which may be attractive if the study-specific parameters $\theta_i$ are only of secondary interest \citep{borenstein2021introduction}.

\subsection{Posterior Distributions}\label{posteriors}
To simplify notation we set $\boldsymbol{y} \coloneqq (y_1,\ldots,y_k)'$ for the observed effect sizes and $\boldsymbol{\sigma^2} \coloneqq (\sigma_1^2,\ldots,\sigma_k^2)'$ for the respective sampling variances, pooled from our $k$ studies. Given an improper uniform or normal prior for the effect $\mu$, the effect's conditional posterior distribution for a given heterogeneity $\tau^2$, $p(\mu \mid \tau^2,\boldsymbol{\sigma^2}, \boldsymbol{y})$, is again normal \citep{JSSv093i06}. In case of an improper uniform prior we obtain $\hat{\mu} = \frac{\sum_{i=1}^k w_i y_i}{\sum_{i=1}^k w_i}$ and $\hat{\sigma} = 1/\sqrt{\sum_{i=1}^k w_i}$ as posterior estimators for the mean and the standard deviation, respectively, where $w_i = (\sigma_i^2 + \tau^2)^{-1}$. In case of a normal prior for $\mu$ with mean $\mu_p$ and variance $\sigma_p^2$ we obtain posterior mean $\hat{\mu} = \frac{\mu_p/\sigma_p^2 + \sum_{i=1}^k w_i y_i}{\sigma_p^{-2} + \sum_{i=1}^k w_i}$ and standard deviation $\hat{\sigma} = 1/\sqrt{\sigma_p^{-2} + \sum_{i=1}^k w_i}$ \citep{gelman2014bayesian,JSSv093i06}.

To additionally estimate study specific means we consider shrinkage estimates.

\subsection{Shrinkage Estimates of study-specific means}\label{shrinkage}

In many applications the main interest focuses on the overall mean estimate of $\mu$. However, here we are interested in estimating the study-specific parameter $\theta_i$, which is informed by both the $i$-th study estimate as well as the estimated overall mean effect and the between-study heterogeneity. This means we are interested in inferring the posterior distribution of the study-specific effect $\theta_i$ for study $i$ with $i = 1,\ldots,k$. In the following we assume the standard NNHM model. Thus, conditional on a heterogeneity variance of $\tau^2$, the posterior distribution of $\theta_i$ is normal with moments given by
\begin{align}
    \mathbb{E}[\theta_i \mid \tau^2, \boldsymbol{\sigma^2}, \boldsymbol{y}] & = \frac{y_i/\sigma_i^2 + \hat{\mu}(\tau^2)/\tau^2}{\sigma_i^{-2}+\tau^{-2}},\label{shrinkage_EW}\\
    Var(\theta_i \mid \tau^2, \boldsymbol{\sigma^2}, \boldsymbol{y}) & = \frac{1}{\sigma_i^{-2}+\tau^{-2}} + \left(\frac{\tau^{-2}}{\sigma_i^{-2}+\tau^{-2}}\hat{\sigma} \right)^2,
\end{align}
where $\hat{\mu}(\tau^2)$ is the estimated summary effect and $\hat{\sigma}$ the estimated standard deviation of $\hat{\mu}$ \citep{gelman2014bayesian}. These formulas illustrate the pull of the $i$th study effect estimate toward the estimated population mean. Hence the term \enquote{shrinkage estimate}. Although the posterior distribution of $\theta_i$ conditional on $\tau^2$ is normal, in practice the amount of heterogeneity $\tau^2$ is unknown and must be estimated. The resulting posterior distribution for $\theta_i$ will be a mixture distribution \citep{JSSv093i06}, which we denote as $F_i$ for study $i$. Two possibilities for constructing confidence intervals (CIs) are (a) define a $(1-\alpha)$-CI via $[c_{i,\alpha/2},c_{i,1-\alpha/2}]$, where $c_{i,\alpha/2}$ is the $\alpha/2$ quantile of $F_i$ or (b) numerically determine a \enquote{shortest} interval, which for unimodal posteriors is equal to the \textit{highest posterior density region} \citep{gelman1995bayesian}. We apply approach (b) using the \texttt{R} package \texttt{bayesmeta}, in order to modify results from a recent study on acute kidney injury in COVID-19 patients by \cite{hardenberg2021critical}.

\section{Acute Kidney Injury and COVID-19}
\label{sec:aki-covid19}
AKI is a relevant complication in COVID-19. Some studies have reported that the incidence of AKI in COVID-19 patients is more than twice the incidence of AKI in non COVID-19 patients \citep{10.3389/fmed.2021.719472}. Furthermore, evidence suggests that patients suffering from AKI experience substantially increased in-hospital mortality compared with those without AKI. However, the underlying relationships are not fully known. In order to shed some light on these mechanisms, an observational cohort study of 223 COVID-19 patients treated at three sites of a tertiary care center in Berlin, Germany at the Charité-Universitätsmedizin Berlin was performed. Both descriptive statistics and multivariable Cox regression models with time-varying covariates were used to identify risks factors of severe AKI \citep{hardenberg2021critical}. The latter considered time-to-severe AKI as the primary endpoint with prior death as a competing risk. Their study will be our main focus in the application of the methodology described in Section \ref{sec:Intro}.

\subsection{Risk factors for AKI development in COVID-19 Patients}

\cite{10.3389/fmed.2021.719472} conducted a systematic review and meta-analysis of (potential) risk factors for AKI in adult patients with COVID-19. They extracted their data based on a systematic literature search in PubMed, EMBASE, Web of Science, the Cochrane Library, CNKI, VIP and WanFang data between December 1, 2019 and January 30, 2021 \citep{10.3389/fmed.2021.719472}. We use their data and results to add to the findings in \cite{hardenberg2021critical}. For this analysis we additionally follow \cite{hardenberg2021critical} with respect to variable selection, i.e. with respect to selecting the main potential risk factors for AKI development. Their results are based on a competing risks model with time-dependent covariates. AKI was diagnosed according to KDIGO guidelines in all considered studies. We will consider the following (potential) risk factors for the development of AKI in COVID-19 patients: obesity (BMI $> 30$), gender, smoking, hypertension, diabetes, mechanical ventilation (MV) and the intake of vasopressors. The latter two were deemed especially important risk factors in \cite{hardenberg2021critical}. For all (binary) risk factors, we compare the AKI prevalence (any of the three stages vs. none) in the corresponding patient subgroups for each study, as in \cite{10.3389/fmed.2021.719472}. We note that this is in contrast to the study by \cite{hardenberg2021critical}, who considered severe AKI, i.e. stage 3 AKI, as primary event of interest. However, the studies collected by \cite{10.3389/fmed.2021.719472} all compared any stage AKI with none. The effect measures are log-odds ratios (ORs).

\subsection{Results}

In the following Figures \ref{fig:mv}--\ref{fig:hypertension} (\ref{fig:smoking}--\ref{fig:gender} in the Appendix) we show forest plots for each risk factor separately. Displayed are the study-specific effects and their shrinkage estimates, as well as their corresponding $95\%$ CIs. In each instance the final data row, with shrinkage estimate highlighted in red, represents the main study of interest by \cite{hardenberg2021critical}. Additionally the estimated population means and prediction intervals are provided along with an estimate of $\tau$, the square root of the between-study heterogeneity variance (also with a CI). We will now consider different potential risk factors for the development of AKI in COVID-19 patients successively. Here, $\hat{\mu}$ describes the meta-analytic summary estimate of the effect of the respective risk factor.

Mechanical ventilation (MV) is a statistically significant potential risk factor with a study specific estimate of 3.488 with a $95\%$ CI of [2.686,4.289] and the shrinkage estimate is 3.271 with a $95\%$ CI of [2.529,4.017]. MV was pointed out as one of the main risk factors for AKI in COVID-19 patients in \cite{hardenberg2021critical}. The $\hat{\mu}$ refers to the estimated mean log-OR across all studies. Meta-analysis yields an estimated log-OR of $\hat{\mu} = 2.215$ and $95\%$ CI for $\mu$ of [1.799,2.630]. The estimated heterogeneity is $\hat{\tau} \approx 0.920$.

The use of vasopressors is also a statistically significant potential risk factor with a study specific estimate of 5.095 with a $95\%$ CI of [4.076,6.113]. The respective shrinkage estimate is 4.665 with a $95\%$ CI of [3.676,5.659]. Like MV, vasopressors were pointed out as one of the main risk factors for AKI in COVID-19 patients in \cite{hardenberg2021critical}, based on a competing risks model with time dependent covariates. Based on a meta-analysis, the estimated log-OR was $\hat{\mu} = 2.295$ with a $95\%$ CI for $\mu$ of [1.558,3.015]. The estimated heterogeneity is $\hat{\tau} \approx 1.250$.

Moreover, hypertension is also a statistically significant risk factor with a study specific estimate for the study by \cite{hardenberg2021critical} of 1.352 and a corresponding $95\%$ CI of [0.795,1.909]. The respective shrinkage estimate is 1.053 with a $95\%$ CI of [0.601,1.519]. Meta-analysis yields an estimated log-OR of $\hat{\mu} = 0.687$ with a $95\%$ CI for $\mu$ of [0.535,0.855]. The estimated heterogeneity is $\hat{\tau} \approx 0.330$.

We highlight the differing amounts of heterogeneity in the models for the respective risk factors. For MV, vasopressors and hypertension, $\tau$ was estimated at 0.920, 1.250 and 0.330 respectively. Therefore, for MV and vasopressors the shrinkage estimates are very close to the study-specific estimates by \cite{hardenberg2021critical}, as shown in Figures \ref{fig:mv} and \ref{fig:vasopressors}. For hypertension, shown in Figure \ref{fig:hypertension}, where the estimated heterogeneity is considerably smaller, the summary estimate of $\mu$ has a much larger weight, leading to a more substantial difference between the shrinkage and study-specific estimate. This also becomes evident when one considers the numerator in expression (\ref{shrinkage_EW}). \cite{rover2021bounds} showed that a study's minimum contribution to its own shrinkage estimate is its fixed-effect weight, i.e. the study weight for $\tau^2=0$ (the fixed-effects model). The study's contribution then increases monotonically, as $\tau^2$ increases. For $\tau^2 \rightarrow \infty$, the shrinkage estimate of the $i$th study converges to the study's effect estimate $y_i$ with $i = 1,\ldots,k$.

For the forest plots showing the results of gender, smoking, diabetes and obesity, we refer to the Appendix. These were non-significant potential risk factors in the competing risks model by \cite{hardenberg2021critical}, see their Figure 1. As a reminder to the reader, the effect sizes in the time-to-event competing risks model by \cite{hardenberg2021critical} are cause-specific hazard ratios. \enquote{The cause‐specific hazard ratio denotes the relative change in the instantaneous rate of the occurrence of the primary event in subjects who are currently event‐free} \citep{austin2017practical}. On the other hand, for the more simple group comparisons we consider here, the effects are log-odds ratios. Therefore, there are some important differences that should be kept in mind. For one, the cause-specific hazard ratios incorporate a time component, which the ORs do not. Furthermore, the ORs are unadjusted effects, as we are dealing with simple group comparisons. The cause-specific hazard ratios however, are adjusted effects as they are part of a model that includes various other covariates. Even though gender, smoking, diabetes and obesity were non-significant risk factors respectively in the competing risks model by \cite{hardenberg2021critical} with regards to the cause-specific hazard ratios, the estimated overall ORs for gender, smoking and diabetes were in fact statistically significant. 

Smoking is a statistically significant potential risk factor with $\hat{\mu} = 0.244$ and a $95\%$ CI for $\mu$ of [0.011,0.557]. The study specific estimate for \cite{hardenberg2021critical} is 0.903 with a $95\%$ CI of [0.189,1.616]. The corresponding shrinkage estimate is 0.441 with a $95\%$ CI of [0.031,1.041]. The estimated heterogeneity is $\hat{\tau} \approx 0.210$.

Diabetes is a statistically significant potential risk factor for AKI in COVID-19 patients with an estimated overall effect of $\hat{\mu}=0.548$ and a 95\% CI for $\mu$ of [0.480,0.610]. The study-specific log odds ratio from \cite{hardenberg2021critical} is 0.694 with a 95\% CI of [0.049,1.340]. The corresponding shrinkage estimate is 0.555 with a 95\% CI of [0.389,0.718]. The estimated heterogeneity is very small with $\hat{\tau} \approx 0.045$.

Overall, gender is a statistically significant potential risk factor for AKI in COVID-19 patients with $\hat{\mu}=0.432$ and a $95\%$ CI for the main effect $\mu$ of [0.304,0.567]. The study-specific data from \cite{hardenberg2021critical} yielded an estimated log odds ratio 0.634 with a corresponding $95\%$ CI of [0.073,1.195]. The respective shrinkage estimate is 0.527 with a $95\%$ CI of [0.134,0.927]. The estimated heterogeneity is $\hat{\tau} \approx 0.270$.

Obesity was a special case, as the study by \cite{hardenberg2021critical} presented a considerable outlier compared with the other considered studies (cf. Figure \ref{fig:obesity}). Even though the overall OR of $\hat{\mu}=0.283$ was not statistically significant with a 95\% CI of [-0.113,0.718], the study-specific estimate of the OR for obesity was 1.955 with a 95\% CI of [1.263,2.648]. The resulting shrinkage estimate is 1.383 with a 95\% CI of [0.567,2.178]. The estimated heterogeneity is $\hat{\tau} \approx 0.520$.

Finally, we present the relative shrinkage weights of the study \cite{hardenberg2021critical} for each potential risk factor respectively in Table \ref{tab:shrinkage_weights}. The shrinkage weights lie between 0.856 for vasopressors and 0.054 for diabetes. Here a value of one would mean the shrinkage estimate is exactly equal to the study-specific estimate and a value of zero would mean the shrinkage estimate is exactly equal to the meta-analytic summary estimate $\hat{\mu}$. We remind the reader that as the heterogeneity variance $\tau^2$ increases (decreases), the shrinkage weight of the target study also increases (decreases).

\begin{table}[h!]
\caption{Shrinkage weights of the \cite{hardenberg2021critical} study for each potential risk factor.}
\centering
    \begin{tabular}{ |l|c|  }
\multicolumn{2}{|c|}{Contribution of \cite{hardenberg2021critical} to shrinkage estimate} \\
\hline
Risk factor & Shrinkage weight \\
\hline
Mechanical Ventilation & 0.837  \\
Vasopressors & 0.856   \\
Hypertension & 0.566 \\
Obesity & 0.685 \\
Diabetes & 0.054 \\
Gender & 0.485 \\
Smoking & 0.328 \\
\hline
\end{tabular}
\label{tab:shrinkage_weights}
\end{table}


\clearpage
\begin{sidewaysfigure}
    \centering
    \includegraphics[width=\linewidth]{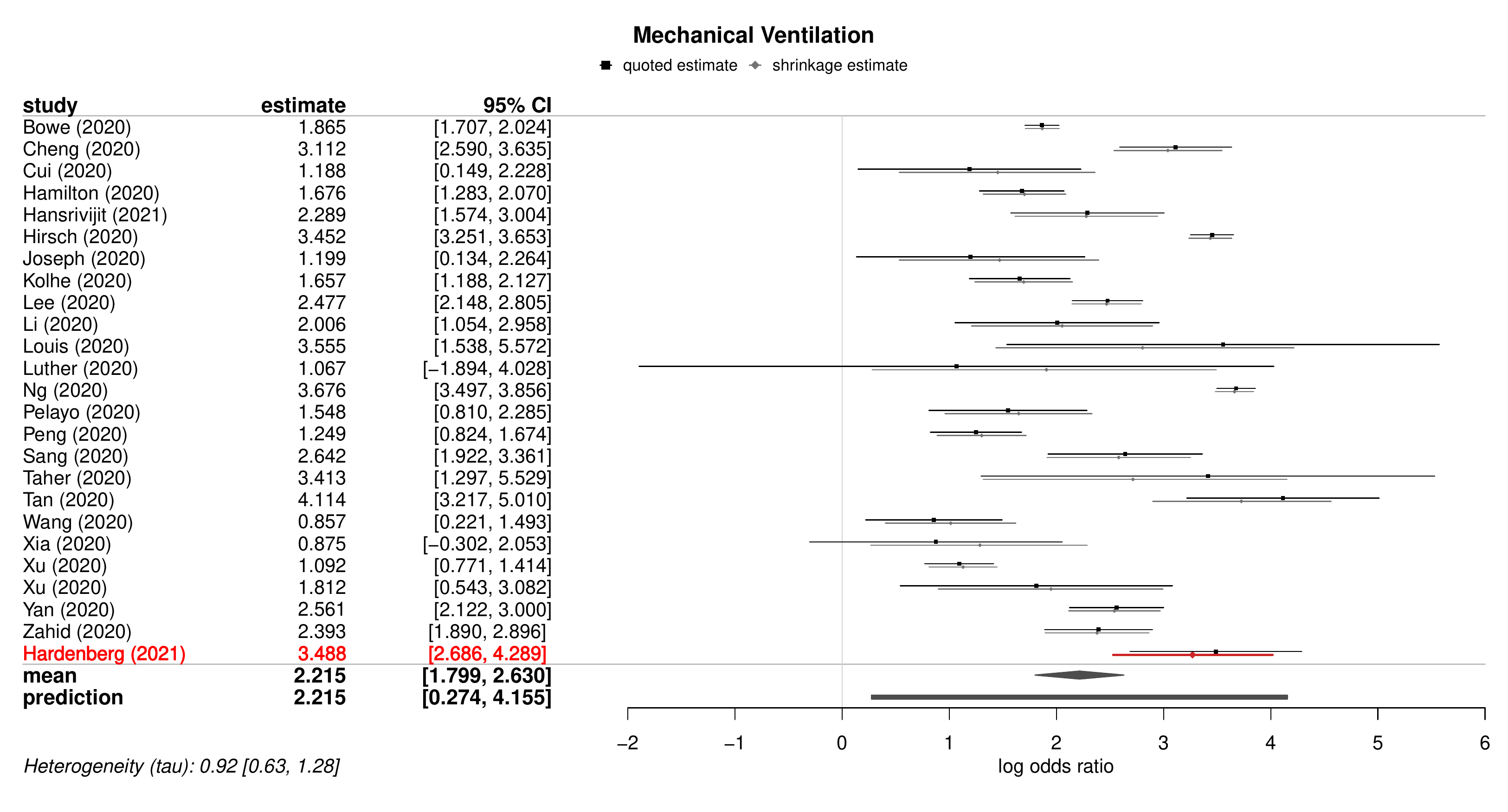}
    \caption{Forest plot and shrinkage estimates for the risk factor mechanical ventilation.}
    \label{fig:mv}
\end{sidewaysfigure}
\clearpage

\begin{sidewaysfigure}
    \centering
    \includegraphics[width=\linewidth]{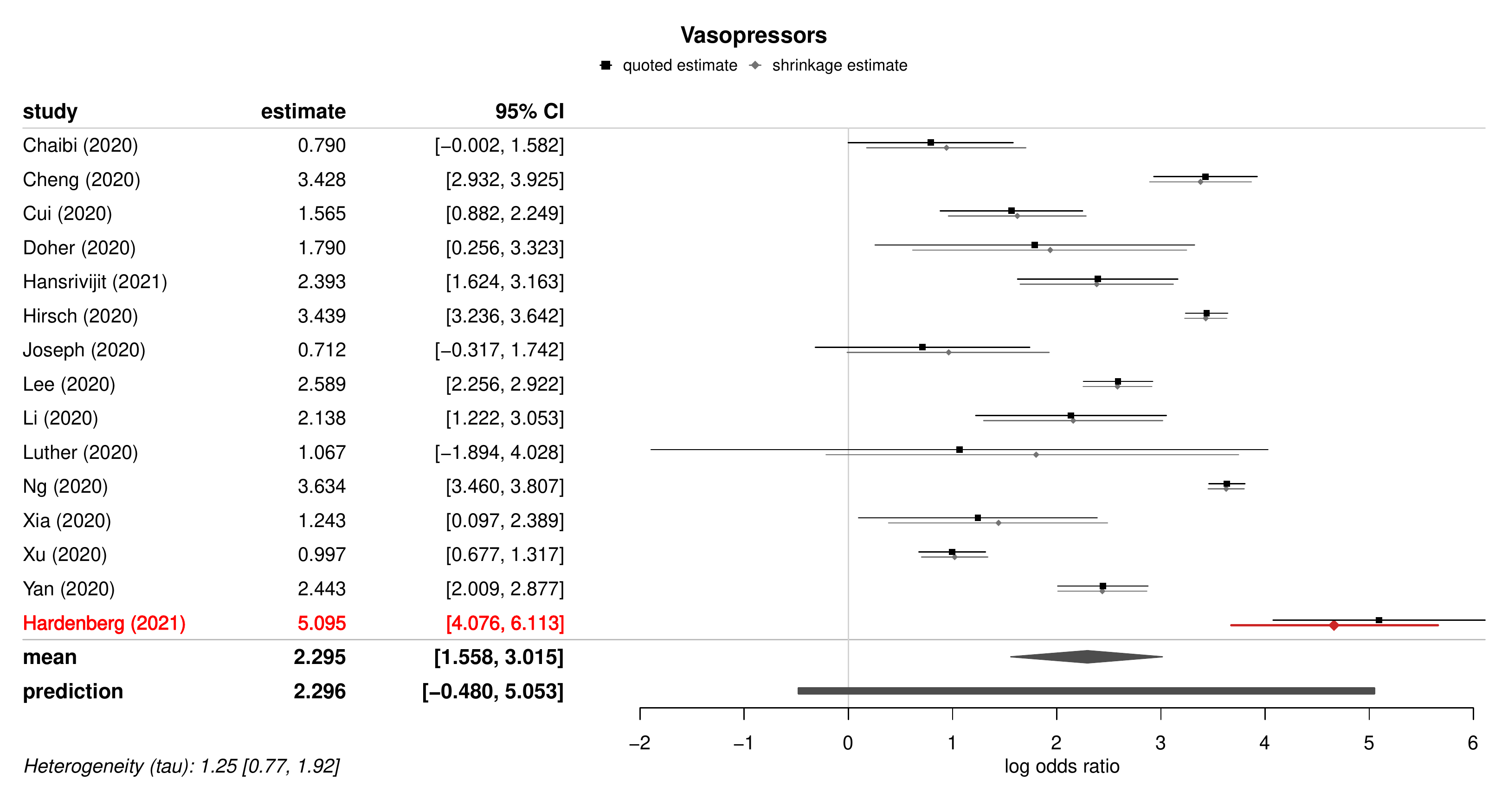}
    \caption{Forest plot and shrinkage estimates for the risk factor vasopressors.}
    \label{fig:vasopressors}
\end{sidewaysfigure}
\clearpage

\begin{sidewaysfigure}
    \centering
    \includegraphics[width=\linewidth]{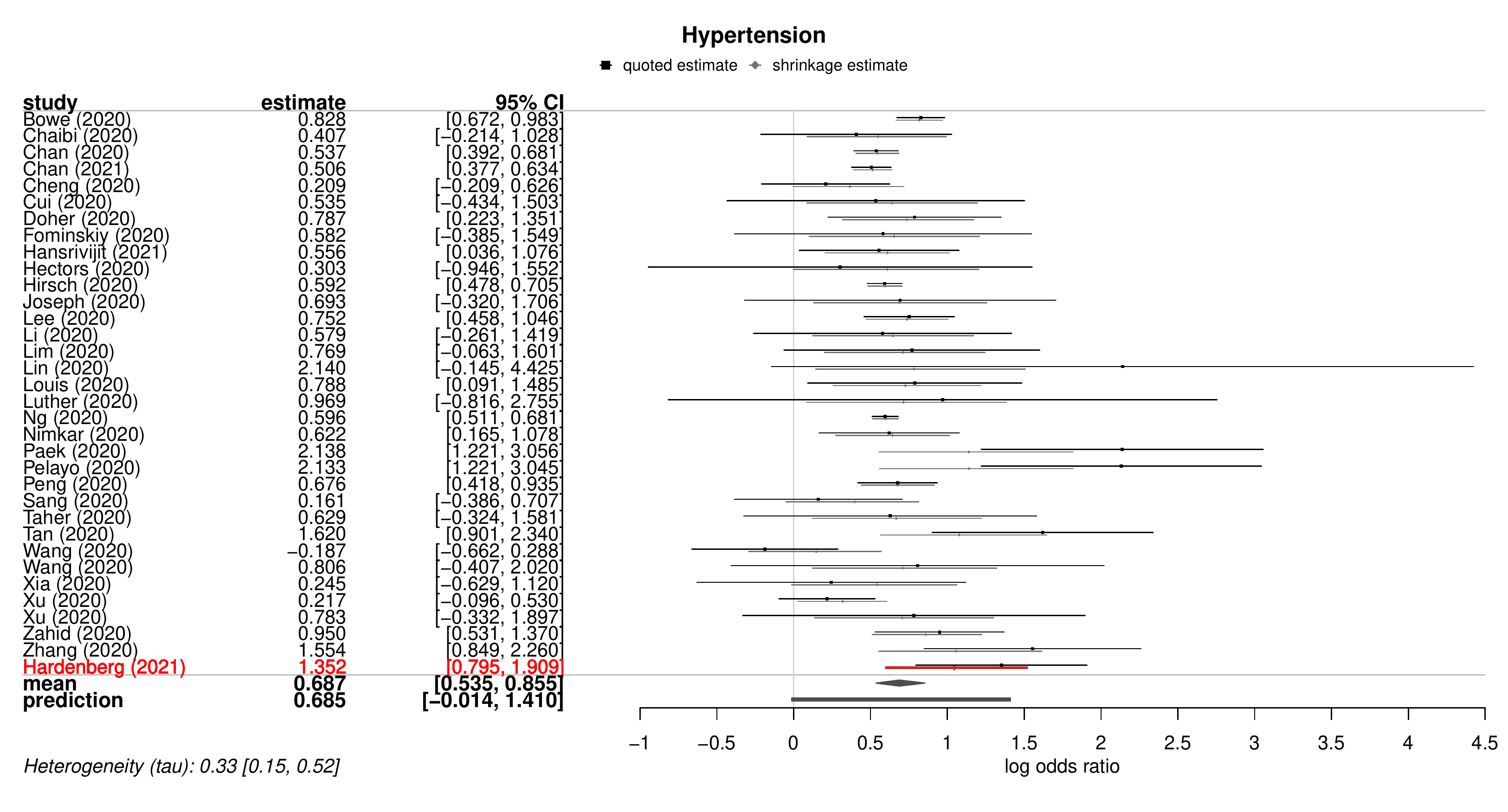}
    \caption{Forest plot and shrinkage estimates for the risk factor hypertension.}
    \label{fig:hypertension}
\end{sidewaysfigure}
\clearpage

\section{Discussion}
\label{sec:discussion}

In this manuscript we demonstrated a Bayesian approach that allows researchers to construct predictive prior distributions based on meta-analyses, yielding posterior distributions and shrinkage estimates for a study of primary interest, which are quantified in the light of external evidence. This approach can for example be easily implemented in \texttt{R} using the \texttt{bayesmeta} package \citep{JSSv093i06}. We apply the discussed approach for a study by \cite{hardenberg2021critical}, who analyze potential risk factors for the development of severe AKI in COVID-19 patients. We expand on their evidence with a meta-analysis on AKI in COVID-19 by \cite{10.3389/fmed.2021.719472}. Although \cite{hardenberg2021critical} focus on time-to-event data and implement competing risks models, due to unavailability of comparable studies, we focus on more simple effects as published by \cite{10.3389/fmed.2021.719472}. The results confirm that both mechanical ventilation and use of vasopressors are important risk factors for the development of AKI in COVID-19 patients. Hypertension also appears to be a risk factor for AKI development that should not be ignored. Furthermore, our analysis of ORs supports that smoking, diabetes and male gender are additional risk factors for AKI development in COVID-19 patients that should be taken into account.

As demonstrated in our analysis, there is a clear connection between the amount of heterogeneity in the meta-analytic model and the amount of borrowing of strength that takes place from external (relative to the primary study of interest) information. This becomes evident, when the relative shrinkage weights for each potential risk factor from Table \ref{tab:shrinkage_weights} are compared with the respective heterogeneity variances. Mechanical ventilation and vasopressors, which have rather large estimated heterogeneity variances with $\hat{\tau}^2$ equal to 0.846 and 1.563 respectively, have the largest shrinkage weights with 0.837 and 0.856. Diabetes on the other hand has a heterogeneity of only $\hat{\tau}^2 \approx 0.002$ with a corresponding shrinkage weight of 0.054.

A limitation of this work is that we only applied the presented Bayesian approach for
the normal-normal hierarchical model as we did not have access to or information on competing risks or covariates in other comparable studies/analyses. Thus, a more sophisticated analysis, e.g. in the context of a meta-regression, was not feasible. The present study was facilitated by having access to the original data from \cite{hardenberg2021critical}, as this allowed us to compute estimates corresponding to those discussed in the meta-analyses by \cite{10.3389/fmed.2021.719472}. Analyses may be more difficult if researchers do not have access to the original data from their primary study of interest.

\bibliographystyle{apalike}
\bibliography{lit} 

\newpage
\section*{Competing interests}
The authors declare that they have no competing interests.

\section*{Author's contributions}
TW performed the data analysis and took charge of manuscript writeup. EK assisted in the research phase and data wrangling. FK was helpful in early discussions and with setting up contacts and obtaining relevant data. JH conducted the main study of focus on acute kidney injury in COVID-19 patients and provided medical expertise. MP provided helpful feedback during conception and manuscript writeup. CR advised on the Bayesian methodology and implementation in \texttt{R}. All authors read and approved the final manuscript.

\section*{Funding}
The work of Christian Röver, Thilo Welz and Markus Pauly was supported by a VW Corona Crisis and Beyond Grant. The funders had no role in data collection, analysis or preparation of the manuscript.

\section*{Availability of data and materials}

The \texttt{R} code will be made publicly available at the repository \texttt{https://osf.io/4rz98/}, pending publication. All relevant data are quoted in figures/tables.

\appendix
\section*{Appendix}

\noindent
\section*{Time-to-acute-kidney-injury-development in COVID-19 patients}
\label{sec:survival}

\begin{figure}[h]
    \centering
    \begin{subfigure}[t]{0.49\textwidth}
        \centering
        \includegraphics[width=\linewidth]{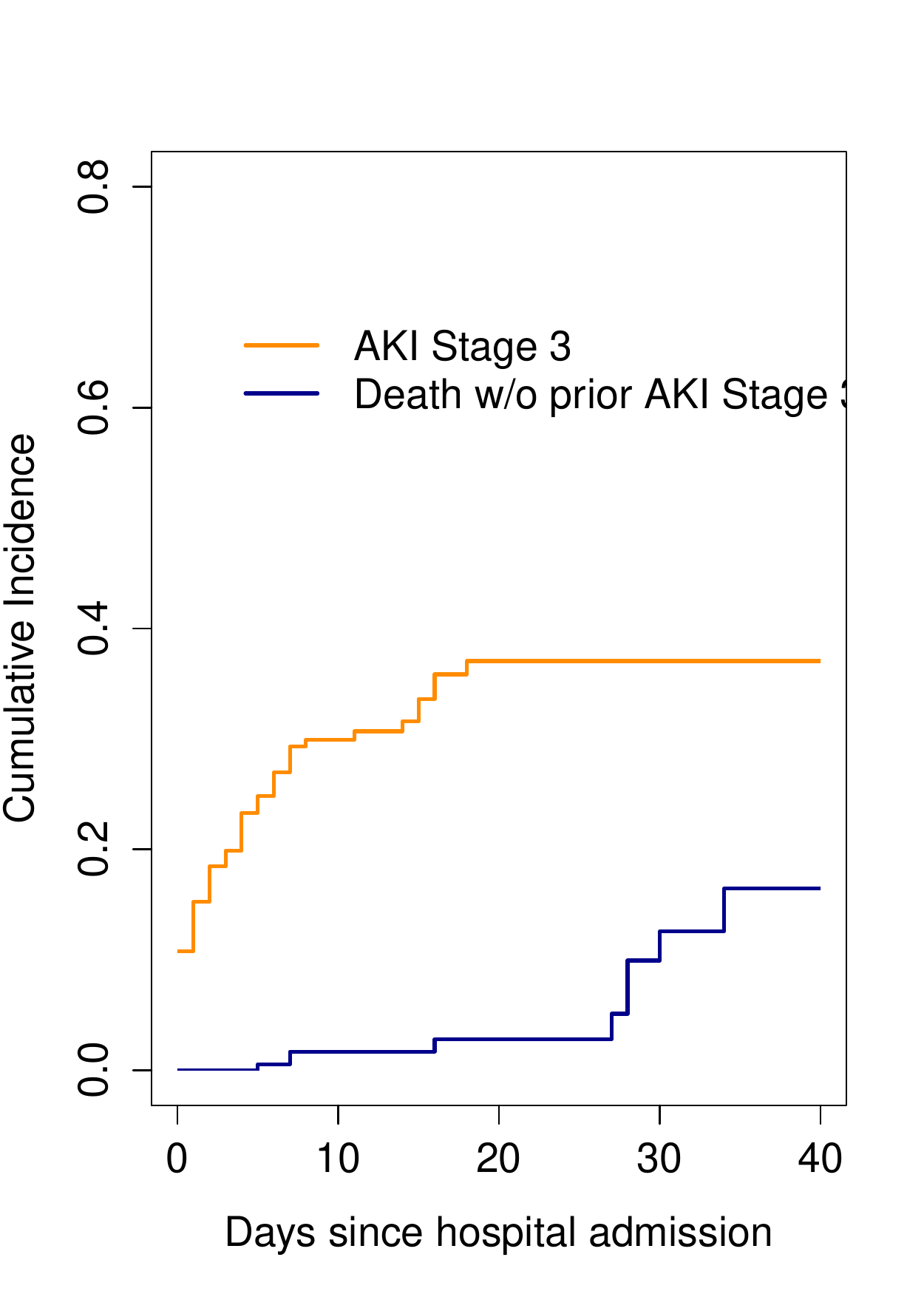} 
        \caption{Cumulative incidence of AKI stage 3 and death without prior AKI stage 3 in COVID-19 patients.} \label{fig:cr}
    \end{subfigure}
    \hfill
    \begin{subfigure}[t]{0.49\textwidth}
        \centering
        \includegraphics[width=\linewidth]{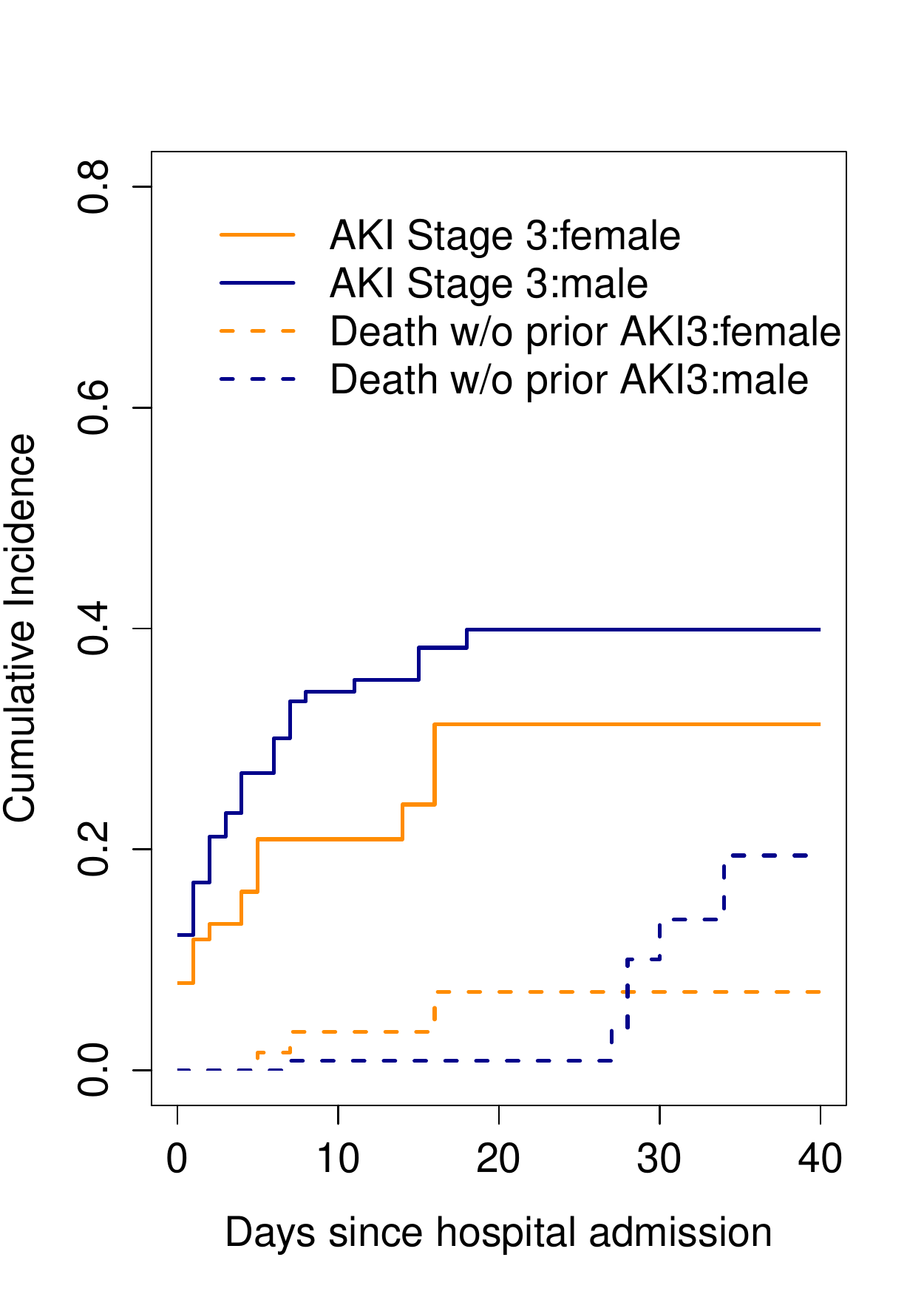} 
        \caption{Cumulative incidence of AKI stage 3 and death without prior AKI stage 3 in COVID-19 patients stratified by gender.} \label{fig:crgender}
    \end{subfigure}
\end{figure}

\cite{hardenberg2021critical} implemented multivariable Cox proportional hazards regression models with time-varying covariates. More specifically, they fit two competing risks models with time-varying covariables: a main model with 11 covariates and a reduced model with 8 covariates. The main event of interest was time to severe AKI (AKI stage 3) with prior death as the competing risk. The respective cumulative incidence functions are shown in Figure \ref{fig:cr}, similar to Supplemental Figure 2 in \cite{hardenberg2021critical}. The results are based on $n=223$ patients, 70 of which experienced severe AKI during hospital admission with 9 patients dying without prior severe AKI (the competing event). Additionally, Figure \ref{fig:crgender} shows the cumulative incidence functions stratified by gender.

\cite{hardenberg2021critical} used these competing risks models to estimate cause-specific hazard ratios for potential risk factors for the development of severe AKI, along with $95\%$ CIs. Their results indicate that MV and a need for vasopressors are especially strong potential risk factors for AKI development in COVID-19 patients. For MV and based on the main model, we obtain an estimated cause-specific hazard ratio of 8.17 with a $95\%$ CI of [1.95,34.25]. For vasopressors, the main model yields a hazard ratio of 3.23 with a $95\%$ CI of [1.28,8.15]. Further notable potential risk factors are a high leucocyte (white blood cells) count, procalcitonin (a precursor of the hormone calcitonin) increase and arguably hypertension. For hypertension the estimated cause-specific hazard ratio was 1.86 with a $95\%$ CI of [0.93,3.72] based on the main model and a hazard ratio of 2.19 with a $95\%$ CI of [1.12,4.29] for the reduced model. The cause-specific hazard for leucocyte counts was estimated at 1.67 for the main model with a $95\%$ CI for [1.00,2.79] and the hazard ratio for procalcitonin was estimated at 1.82 with a $95\%$ CI of [1.11,2.97]. Data regarding leucocyte counts and procalcitonin were not available in the meta-analyses by \cite{10.3389/fmed.2021.719472}. The results above are summarized in Table \ref{tab:hazards}.

 \begin{table}[h!]
 \caption{Cause-specific hazard ratios and confidence intervals based on a competing risks model from \cite{hardenberg2021critical}.}
        \begin{tabular}{cS[table-format = 1.1]
                @{}S[table-format = -1.2]@{,\,}S[table-format = -1.2]@{\,]}
                }
                Risk Factor & \multicolumn{1}{c@{\quad\space}}{Hazard Ratio} & \multicolumn{2}{c}{95\% CI} \\
                \hline
            Mechanical Ventilation & 8.17 & [1.95  & 34.25 \\
            Vasopressors & 3.23 & [1.28 & 8.15 \\
            Hypertension & 1.86 & [0.93  & 3.27 \\
            Leucocyte count & 1.67 & [1.00 & 2.79 \\
            Procalcitonin & 1.82 & [1.11 & 2.97 \\
        \end{tabular}
        \label{tab:hazards}
 \end{table}

In Section \ref{sec:aki-covid19}, we considered the potential impacts of mechanical ventilation, vasopressors and hypertension on AKI development from an alternative viewpoint, i.e. non time-to-event data. There we analyzed simple log odds ratios comparing the occurrence of AKI in respective subgroups of COVID-19 patients. In this chapter we consider time-to-event data and cause-specific hazard ratios in the context of competing risks models. In both settings, MV, vasopressors and hypertension are statistically significant and meaningful (based on effect size) risk factors for AKI development in COVID-19 patients. The log odds ratios tell us that there are more AKI injury cases in those patient subgroups that received MV, vasopressors or that have hypertension compared to those that did not. The cause-specific hazard ratios calculated from the competing risks models additionally show that MV, vasopressors and the presence of hypertension increase the hazard of severe AKI in COVID-19 patients.

Unfortunately we were unable to find other studies that analyzed a Cox proportional hazards regression with competing risks that consider time-to-AKI in COVID-19 patients with prior death as a competing event. Therefore we could not apply the Bayesian approach outlined in Section \ref{sec:Intro} for the competing risks setting.

\begin{sidewaysfigure}[h]
    \centering
    \includegraphics[width=\linewidth]{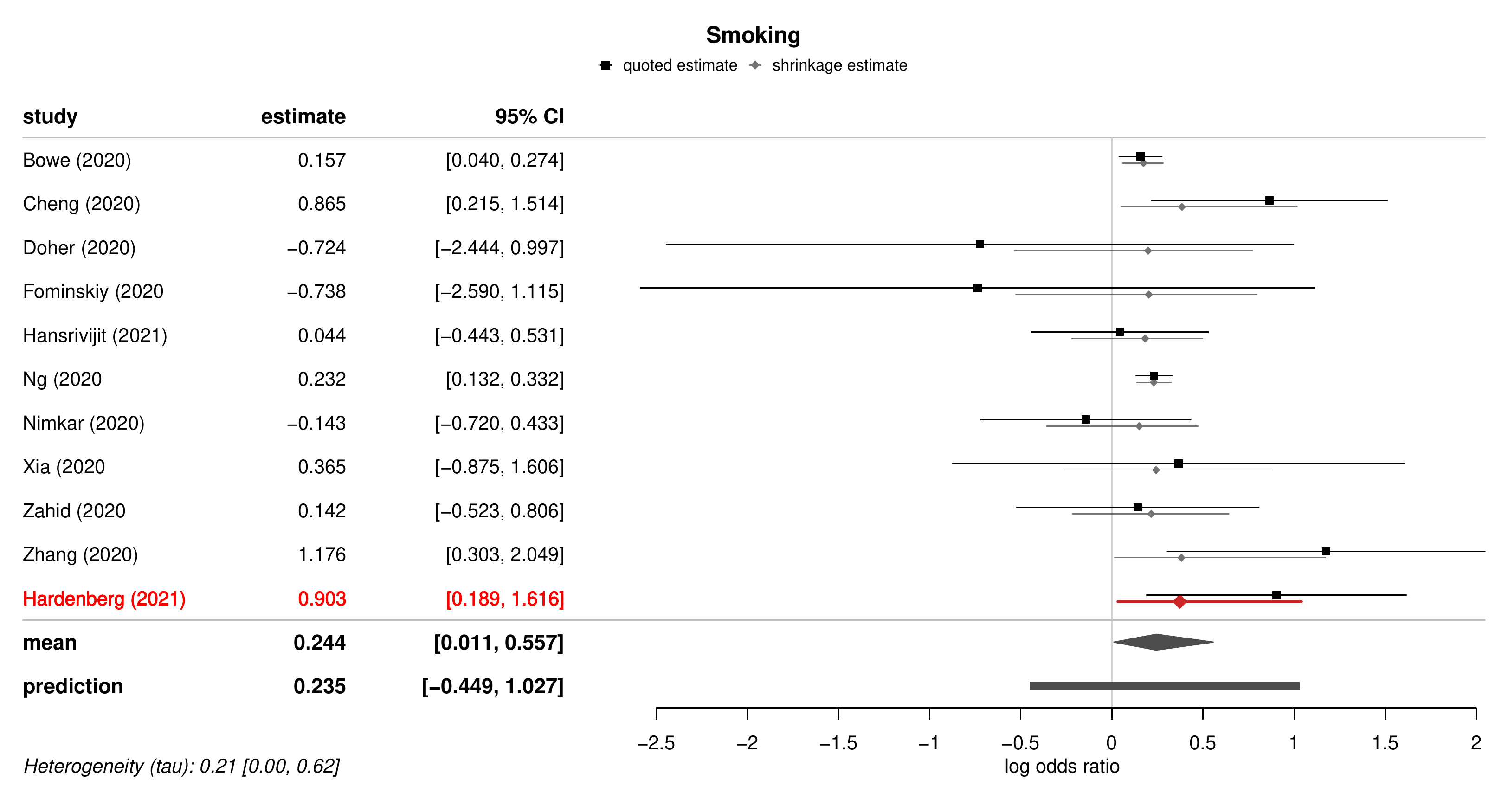}
    \caption{Forest plot and shrinkage estimates for the risk factor smoking.}
    \label{fig:smoking}
\end{sidewaysfigure}

\newpage
\begin{sidewaysfigure}
    \centering
    \includegraphics[width=\linewidth]{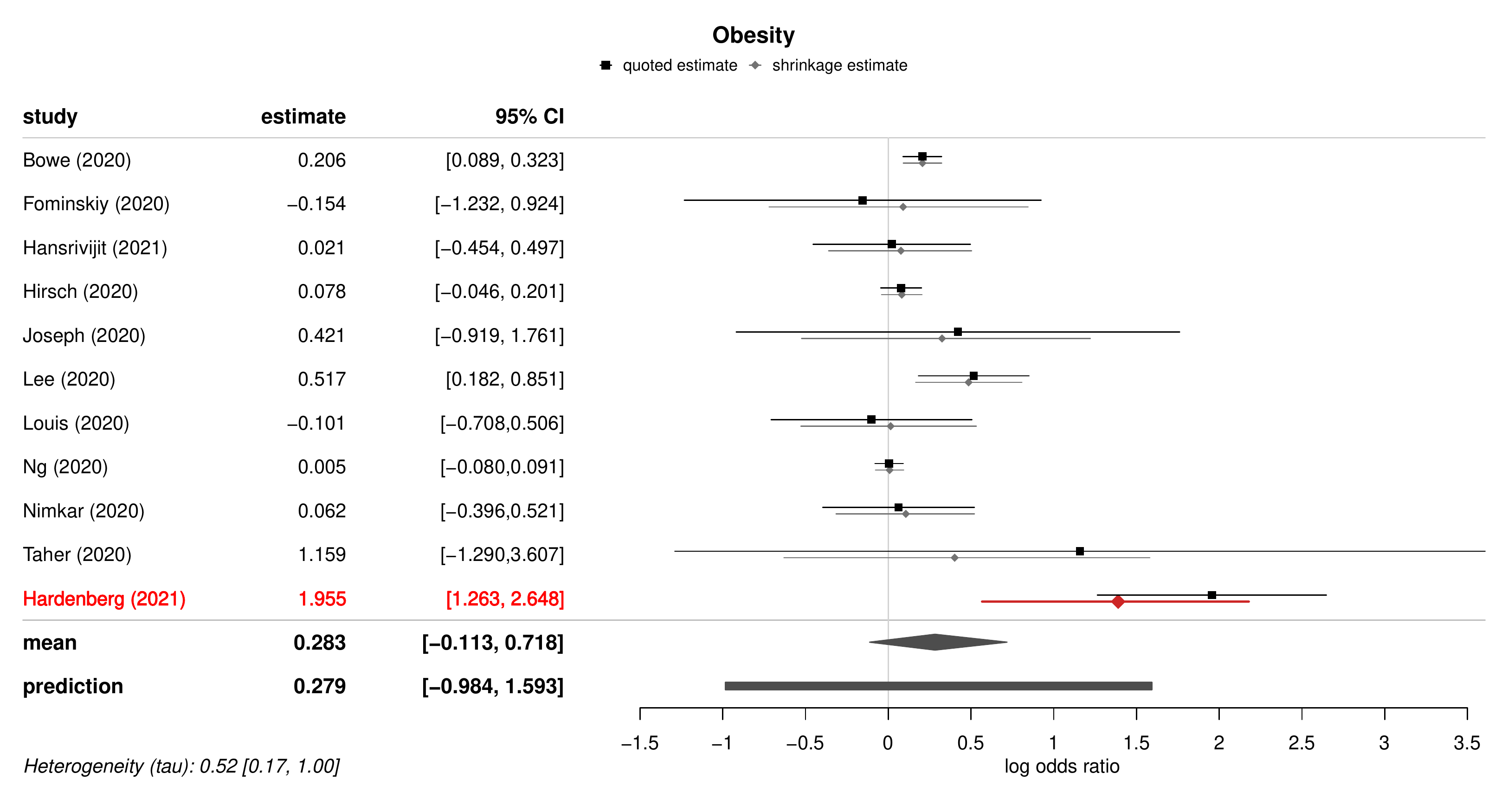}
    \caption{Forest plot and shrinkage estimates for the risk factor obesity.}
    \label{fig:obesity}
\end{sidewaysfigure}

\begin{sidewaysfigure}
\centering
    \includegraphics[width=\linewidth]{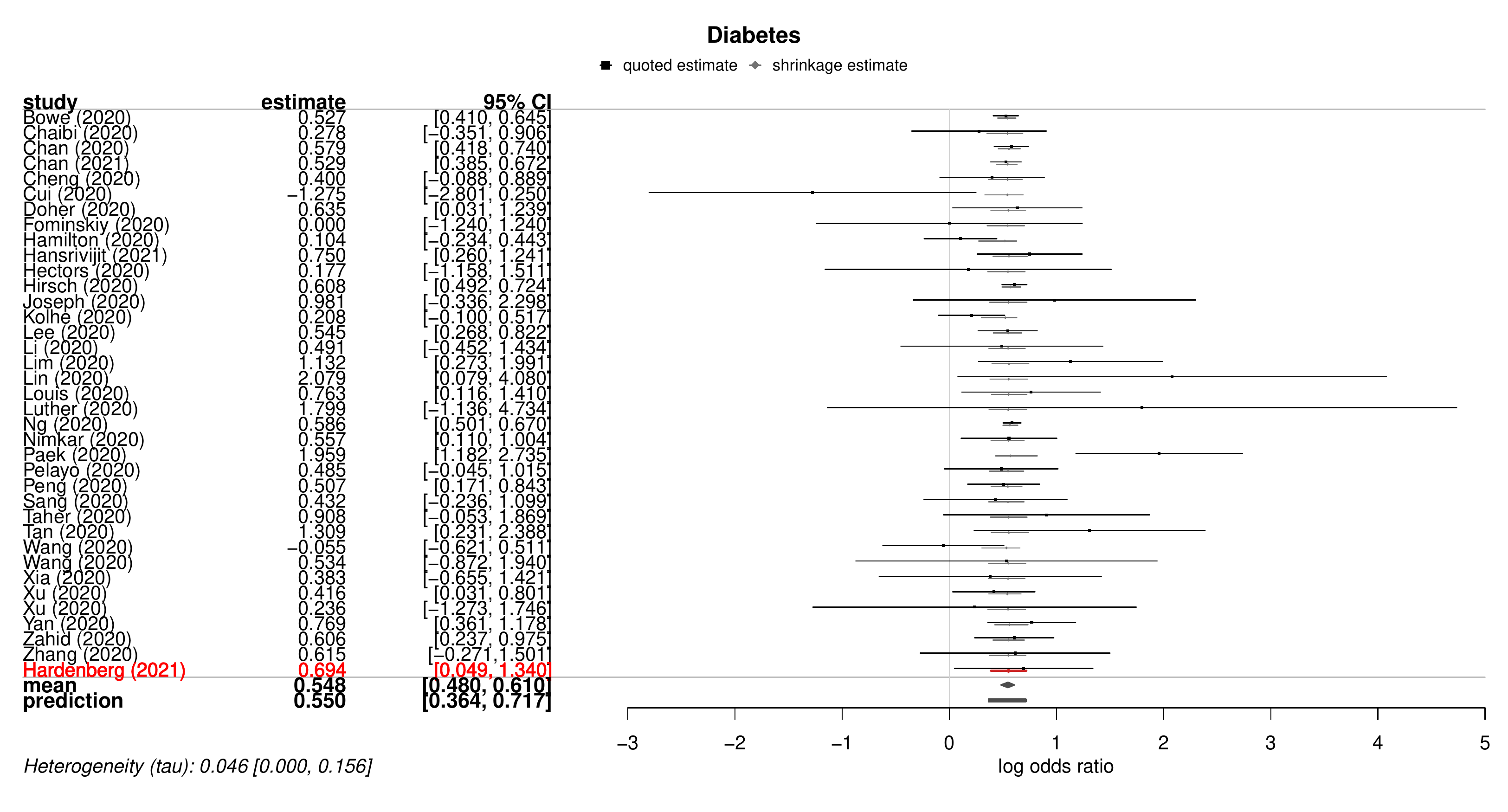}
    \caption{Forest plot and shrinkage estimates for the risk factor diabetes.}
    \label{fig:diabetes}
\end{sidewaysfigure}

\begin{sidewaysfigure}
    \centering
    \includegraphics[width=\linewidth]{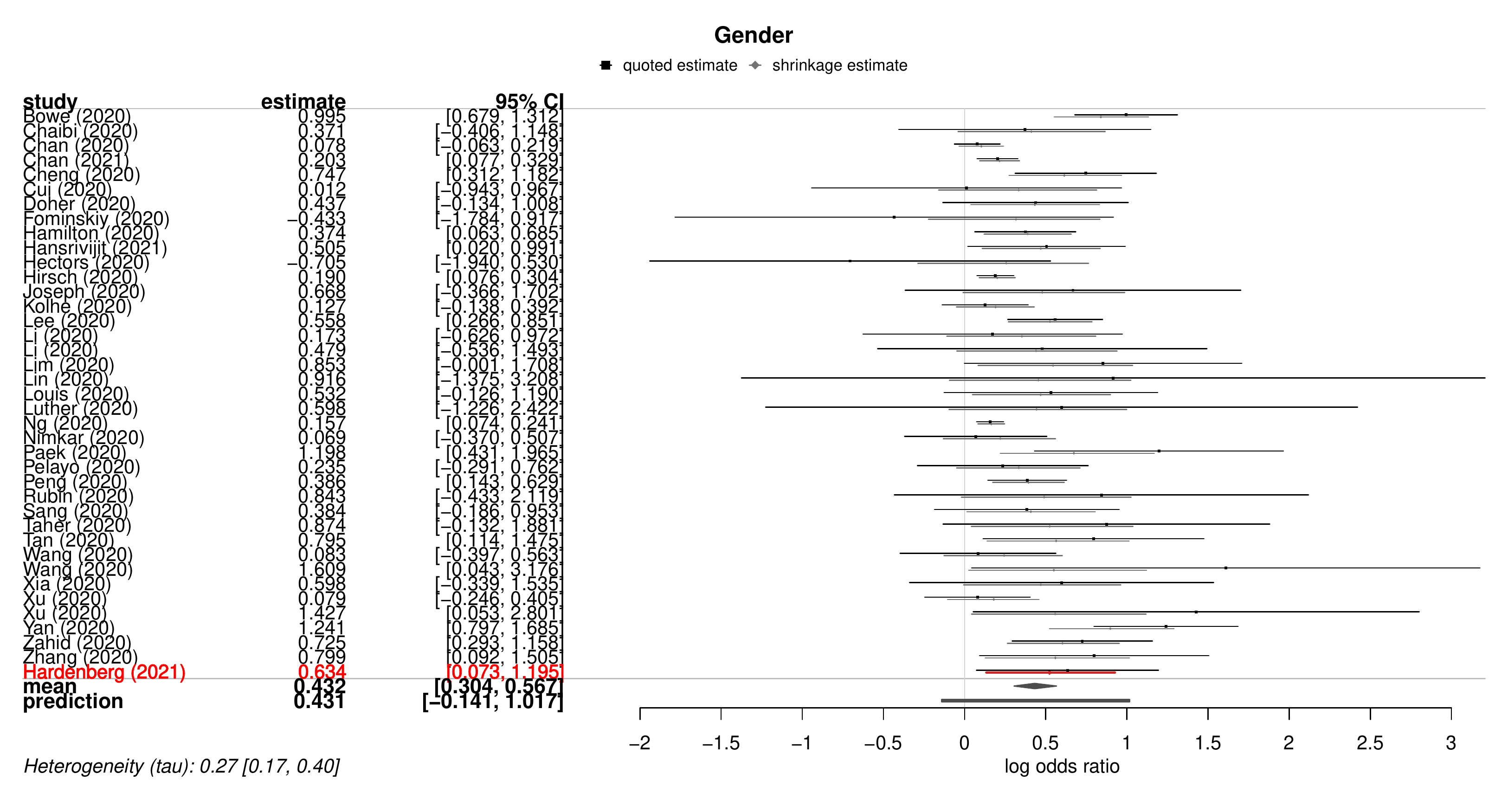}
    \caption{Forest plot and shrinkage estimates for the risk factor gender.}
    \label{fig:gender}
\end{sidewaysfigure}

\end{document}